\documentclass[]{interact}
\usepackage{epstopdf}

\usepackage{natbib}
\bibpunct[, ]{(}{)}{;}{a}{}{,}
\renewcommand\bibfont{\fontsize{10}{12}\selectfont}
\usepackage{csquotes}
\usepackage{tabularx}
\usepackage{caption}
\usepackage[super]{nth}
\usepackage[acronym]{glossaries}
\usepackage[hyphenbreaks]{breakurl}
\usepackage[hyphens]{url}
\usepackage[sf,normalsize]{subfigure}
\usepackage{placeins}
\usepackage[super]{nth}
\usepackage{amsmath, array, makecell}
\usepackage{cellspace} %
\usepackage{changepage}
\usepackage[dvipsnames]{xcolor}

\theoremstyle{plain}

\theoremstyle{definition}

\theoremstyle{remark}

\begin{document}

\articletype{Computer Methods in Biomechanics and Biomedical Engineering: Imaging \& Visualization}

\title{Fast and Robust Femur Segmentation from Computed Tomography Images for Patient-Specific Hip Fracture Risk Screening}

\author{
\name{Pall Asgeir Bjornsson\textsuperscript{a}, Alexander Baker\textsuperscript{b}, Ingmar Fleps\textsuperscript{b}, Yves Pauchard\textsuperscript{c}, Halldor Palsson\textsuperscript{d}, Stephen J. Ferguson\textsuperscript{b}, Sigurdur Sigurdsson\textsuperscript{e}, Vilmundur Gudnason\textsuperscript{e,f}, Benedikt Helgason\textsuperscript{b}, and Lotta Maria Ellingsen\textsuperscript{a,g}\thanks{CONTACT L.M. Ellingsen. Email: lotta@hi.is}}
\affil{\textsuperscript{a}The Dept. of Electrical and Computer Engineering, The University of Iceland, Reykjavik, Iceland; \textsuperscript{b}The Institute for Biomechanics, ETH Zurich, Zurich, Switzerland;\textsuperscript{c}The University of Calgary, McCaig Institute for Bone and Joint Health, Calgary, AB, Canada;\textsuperscript{d}The Dept. of Industrial Engineering, Mechanical Engineering, and Computer Science, The University of Iceland, Reykjavik, Iceland;\textsuperscript{e}The Icelandic Heart Association, Kopavogur, Iceland;\textsuperscript{f}The Dept. of Medicine, The University of Iceland, Reykjavik, Iceland;\textsuperscript{g}The Dept. of Electrical and Computer Engineering, The Johns Hopkins University, Baltimore, MD 21218, USA}
}

\maketitle

\begin{abstract}
Osteoporosis is a common bone disease that increases the risk of bone fracture. Hip-fracture risk screening methods based on ﬁnite element analysis depend on segmented computed tomography (CT) images; however, current femur segmentation methods require manual delineations of large data sets. Here we propose a deep neural network for fully automated, accurate, and fast segmentation of the proximal femur from CT. Evaluation on a set of 1147 proximal femurs with ground truth segmentations demonstrates that our method is apt for hip-fracture risk screening, bringing us one step closer to a clinically viable option for screening at-risk patients for hip-fracture susceptibility. 
\end{abstract}


\section{Introduction}
\label{sec:introduction}
According to the United Nations, more than 40\% of the population of some developed countries will be above the age of 60 by the year 2050 \citep{UN}. This raises concerns about the burden placed on health care systems, since aging societies are associated with a higher prevalence of chronic diseases. 
Policy-makers are thus forced to reconsider the status quo of health care systems, moving away from face-to-face consultation-based care towards a decentralized community or home-based care, as well as transitioning from focusing on treatment to focusing on prevention. 

One of the prevalent chronic diseases suffered by elderly populations is osteoporosis - a bone disease characterized by low bone mass and structural deterioration of bone tissue, leading to bone fragility and an increased risk of fracture.  There is sufficient evidence that the majority of hip fractures are the result of a low trauma fall \citep{hayes,parkkari}. The fracture risk increases with age and, compared to other fracture types, hip fractures are associated with the most dire socioeconomic consequences. 
The most abysmal, and perhaps surprising, statistic is that 11-23\% of individuals will be deceased six months after incurring the fracture, increasing to 22-29\% after one year has passed since the incident \citep{haleem}.

\subsection{Current Standard in Screening for Hip Fracture Risk}
The present clinical ``gold standard'' to diagnose osteoporosis is the areal bone mineral density (aBMD) derived from dual-energy X-ray absorptiometry (DXA). However, a shortcoming of this method is that  even though low aBMD scores are associated with population-based fracture risk, between 36-72\% of incident fractures are sustained by individuals who do not have osteoporosis \citep{stone, schuit, wainwright}. Moreover, the aBMD lacks specificity when stratifying risk considering the fact that the majority of subjects with osteoporosis do not incur hip fractures in their lifetime.

\subsection{Finite Element Analysis} \label{sec:fe}
In order to improve both the specificity and sensitivity of hip fracture screening, X-ray computed tomography (CT) image-based, subject-specific finite element (FE) models of the proximal femur have garnered significant attention and shown promise as a means to overcome the limitations of assessing hip fracture risk using aBMD.
The motivation for this application is to incorporate it into a clinical screening tool that uses FE analysis for hip fracture risk prediction. The widespread use of such tools has the potential to dramatically reduce the economic toll of hip fractures on our healthcare systems, as well as mitigate the potentially devastating consequences for patients. Thus far, several hindrances have impeded clinical translation of FE analysis for hip fracture risk screening: the cost of hiring trained engineers to carry out simulations instead of the clinical staff, the ambiguous accuracy of these methods for fracture prediction, and the health risk of X-ray exposure caused by the CT scanner. In order to bring clinical applications to fruition, a robust and automated workflow for constructing the FE model and subsequent analysis is imperative. Fleps et al. \citep{Ingmar} demonstrated that femoral strength based on finite element analysis (FEA) can improve hip fracture risk assessment and we employed the same FE pipeline in this work. The workflow pipeline entails the segmentation of the CT image data, generating an FE mesh, applying heterogeneous gray level based material properties to the FEs, applying boundary conditions, solving FE equations and processing the results \citep{yves}.  The aim of this work is to develop a fully automated segmentation of the proximal femur from a CT image, without the need for any manual intervention during postprocessing. 

\subsection{Related Work}
\begin{figure}[h!]
\centering
\includegraphics[width=.6\columnwidth]{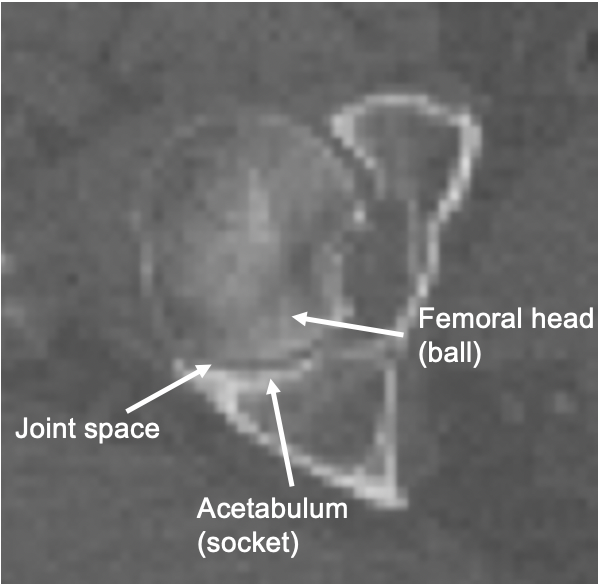}
\caption{An axial CT image slice showing the distinction that our model has to make in order to correctly segment the femoral head.}
\label{acetabulum}
\end{figure}

Bone segmentation of CT images is an elusive problem for several reasons. Firstly, there is an overlap of the Hounsfield units (HU)\footnote{The Hounsfield unit is a relative quantitative measurement of radio density used by radiologists in the interpretation of CT images.} of the bones and surrounding tissue, rendering it impossible to segment solely based on intensity value. Moreover, bones themselves do not have uniform densities, nor do certain bone diseases affect all parts of the bone in the same manner. 
Adjacent bones pose an additional problem when the joint space approaches the resolution of clinical CT data, which is often the case for elderly subjects, and can result in poor segmentations. 
Hence, a method is needed to detect and connect thin and diffusive bone structure boundaries to obtain acceptable segmentations. Figure \ref{acetabulum} displays an example of a hip joint from the data set at hand, which can prove challenging to segment if the boundary between the femoral head and acetabulum is unclear.

Promising methods for segmenting the proximal femur from CT images that have gained traction as of late are statistical shape models (SSM) \citep{chang, younes}, multi-atlas segmentation \citep{wang,chu} and graph-cut segmentation \citep{yves}. The two former methods, however, require a database of gold standard segmentations, while the latter method does not necessitate such prior knowledge. Lastly, the implementation of 3-dimensional (3D) convolutional neural networks (CNNs) to address the problem of femur segmentation is one of the most recent developments \citep{chen, zhao} and has become the method of choice in biomedical image analysis. 

One of the most successful previous methods for segmenting the proximal femur is the aforementioned graph-cut method, carried out by Pauchard et al. \citep{yves}. 
In short, this method separates the background class from the target object by finding the global minimum of a cost function. If differences in intensity are large (i.e., at object boundaries) with respect to $\sigma^{2}$ (variance of homogenous regions in the image), then the cost of cutting an edge is low. On the other hand, if differences are small in comparison to $\sigma^{2}$, then the cost is high. 
They reported a mean Dice Similarity Coefficient (DSC) \citep{dice} of $0.973 \pm 0.005$, while the mean Hausdorff distance (HD) between manual segmentations and interactive graph-cut segmentations was $3.75 \pm 1.26$mm. This method, however, suffers from its only partially autonomous nature: when producing the segmentation predictions, manual input is required by the user to initiate the graph-cut segmentation process, which is a key limitation of this method.  

Zhao et al. \citep{zhao} proposed an automated, patch-based 3D v-net architecture \citep{v-net} (employing the Dice loss function) on a cohort that comprised 397 quantitative computed tomography (QCT) scans, of which only 10\% was used to evaluate the model. This reliance on such a large training/validation set, which is not always available, is a key limitation of this model. 
The method struggled to segment the femur around the most dynamic sections (i.e., the femoral head), resulting in some unacceptable segmentations. Nevertheless, the authors reported a mean DSC of $0.9815\pm 0.0009$ and, for a subject with 60 QCT slices, a segmentation time of 15s.

Another automated 3D CNN method conducted by Chen et al. \citep{chen} is based on the u-net architecture \citep{unet} and employs both the Dice loss and the Jaccard loss functions to segment the entire femur. An edge detection task was embedded into a fully convolutional network (FCN) to address the problems of diffusive joint spaces and weak femur boundaries. 
The method, however, shares the same limiting factor as the previously discussed method \citep{zhao} in that it requires a large training set (120 samples) which, for many biomedical segmentation tasks, is not a viable option. The authors of this study reported a mean DSC of $0.9688\pm 0.0095$ on an evaluation set of 30 CT images. 
Table \ref{prior_work_table} compares the methods mentioned in this section. 

The key limitation to previous deep learning methods has been the reliance on a vast training set, which requires an equally large set of ground truth segmentations. The proposed method necessitates far fewer ground truth segmentations than these methods. A preliminary version of our model was reported in conference form ~\citep{bjornsson}; here the method has been validated on a signiﬁcantly larger test set and compared with a state-of-the-art segmentation method. Moreover, we demonstrate its use in our patient-specific screening method, where the FEA-derived femoral strength based on our method was compared to that based on ground truth segmentations, further validating the method on the end-product. 
\begin{table*}[t]
\centering
\begin{adjustwidth}{-2.5cm}{}
\caption{A comparison between different femur segmentation methods. Here, $\sigma$ denotes standard deviation. The time required to segment a femur volume should be interpreted with caution considering the varying workstations, resolution, and number of slices per scan.}
     \sffamily \begin{tabularx}{\linewidth}{p{2.5cm}  p{2cm} p{1.75cm} p{1.25cm} p{2.25cm} p{2.25cm} p{2cm}}
    \hline
   \textbf{Authors} \hfill & \textbf{Model} \hfill & \textbf{Training set (CNNs only)}  \hfill & \textbf{Test set}  \hfill & \textbf{Slice thickness [mm]} & $\pmb{\mathrm{DSC}\pm \sigma}$ & \textbf{Time p. femur (s)}\\ \hline
    Pauchard et al. \cite{yves} & Graph-cut & N/A & 48 CT & 1 & $0.973\pm 0.005$ & 120-300 \\
    Chu et al. \cite{chu} & Multi-atlas & N/A & 30 CT & 1 & $0.979\pm0.029$ & 275 \\
    Younes et al. \cite{younes} & SSM & N/A & 8 CT & - & $0.89 \pm0.01$ & 180 \\
    Chang et al. \cite{chang} & Conditional random field & N/A & 60 CT & 0.45-1.2 & $0.973\pm0.0095$ & - \\
    Chen et al. \cite{chen} & 3D CNN & 150 CT & 30 CT & 1.32-1.85 & $0.9688\pm 0.0095$ & 56 \\
    Zhao et al. \cite{zhao} & 3D CNN & 357 QCT & 40 QCT & 3 & $0.9815\pm 0.0009$ & 15 \\ \hline
    \end{tabularx} 
    \label{prior_work_table}
    \end{adjustwidth} 
\normalfont
\end{table*}

\subsection{Rationale for Deep Learning Approach}
Current femur segmentation methods mostly require a ``user-in-the-loop'' paradigm in order to manually correct segmentations and produce acceptable masks for FE modeling. This lack of robustness is costly in terms of time and the need for highly trained specialists to manually correct the segmentation predictions. Consequently, these methods cannot process larger cohorts to the same degree as a fully automated one, rendering them impractical for clinical application. 
The justification for using a DNN is almost entirely a byproduct of the u-net. Since large data sets containing CT images from a particular scanner are hard to come by, neural networks were not viewed as a particularly attractive alternative for application in biomedical imaging. However, the u-net architecture proposed by Ronneberger et al. \citep{unet} demonstrated fast and precise segmentation without the need for a large data set. DNNs have, as a result, become the state-of-the-art method for segmentation in biomedical image classification \citep{shao, huo}. Instead of requiring minutes to generate a segmentation prediction, CNNs can produce an output in the matter of seconds.

\subsection{Contributions of our Work}
In this paper, we propose a robust, fully automated, and fast segmentation of the proximal femur from CT images. The most salient contributions of our work to the field of biomedical image segmentation of the proximal femur are the following: First, our method takes a human-out-of-the-loop approach rendering the arduous and time-consuming task of making ad hoc corrections unnecessary; second, the model is highly robust, and hence, opens up the possibility of e.g. low-cost opportunistic screening for hip fracture risk based on existing CT data; and third, the processing time (in a matter of seconds) is well within reasonable bounds for clinical implementation.

\section{Materials and Methods} \label{method}
\subsection{The AGES-RS Cohort}
The Icelandic Heart Association (IHA) provided us with CT scans from the Age, Gene/Environment Susceptibility-Reykjavik Study (AGES-RS) \citep{ages}: a cohort that consists of both men and women born between 1907 and 1935, monitored in Iceland by the IHA since 1967. This unique database of high-quality CT images contains roughly 4800 density calibrated CT scans of the proximal femur at baseline and 3300 scans of the same individuals acquired at a five-year follow-up. The resolution of each scan is $512\times512$ voxels with $0.977 \times 0.977 \times 1~\mathrm{mm}^3$ voxel size and the number of slices ranges from 88 to 178. Our model was evaluated using two subsets of the AGES data set. The first subset (Sample I\footnote{This sample set was used by Pauchard et al. \citep{yves} to evaluate their model.}) comprises 48 ``gold standard'' manually delineated proximal femur segmentations from 24 CT images. The second subset within the AGES data set (Sample II\footnote{This sub-cohort from AGES-RS, including all fracture cases, was used by Enns-Bray et al. \citep{ws_enns}.}) consists of 1207 manually delineated segmentations, generated with a semi-automated delineation protocol, that served as ground truth annotations. The proposed segmentation model was trained w.r.t. 60 of these ground truth segmentations and evaluated on the remaining 1147.

\subsection{Validation and Loss Function} \label{validation_and_loss_function}
Since the femur only makes up a small part of each slice, there is a class imbalance problem that must be addressed to avoid the more prevalent class from dominating. The learning process tends to get trapped in a local minima of the loss function and yields a network whose segmentation predictions are heavily biased towards the background class. To combat this problem, the DSC \citep{dice}, which effectively renders the relative spatial areas of each class irrelevant, was implemented.  The DSC measures spatial overlap between segmentations and is given by the following equation:
\begin{equation} \label{eq:dice}
    \begin{aligned}
    \mathrm{DSC} = \frac{2 \sum_{i}^{N} p_i g_i}{\sum_{i}^{N} p_{i}^{2} + \sum_{i}^{N} g_{i}^{2}}.
    \end{aligned}
\end{equation}
In this equation, $N$ is the number of voxels of the predicted binary segmentation volume $p_i \in P$ and the ground truth binary volume $g_i \in G$ \citep{v-net}.
The values of the DSC are restricted to the range $[0,1]$, where $\mathrm{DSC} = 0$ indicates total misclassification and $\mathrm{DSC} = 1$ indicates perfect classification. In order to formulate a loss function, the Dice loss function is defined as $1 - \mathrm{DSC}$.

\subsection{The Proposed Segmentation Pipeline}
\begin{figure*}[t]
\centerline{\includegraphics[width=.9\textwidth]{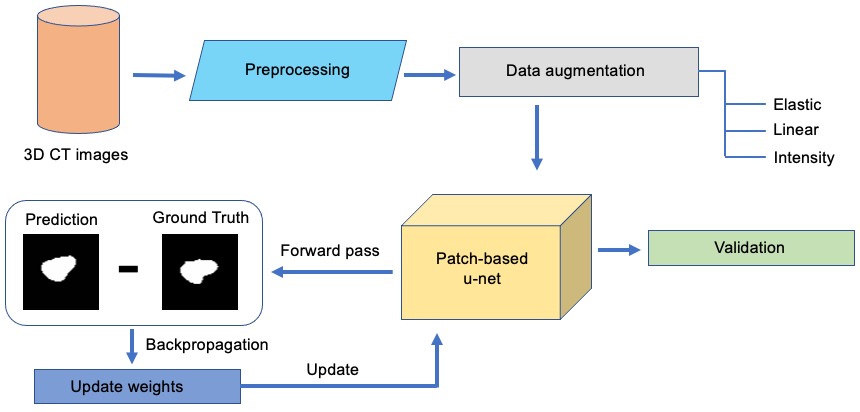}}
\caption{A flowchart showing the workflow of our proposed method.}
\label{workflow}
\end{figure*}

The implemented fully automated proximal femur segmentation pipeline is illustrated in Figure~\ref{workflow} and consists of the following components:
\begin{itemize}
    \item A training/validation set of 30 3D CT images (i.e., 60 proximal femurs) from Sample II of the AGES cohort \citep{ages}
    \item Normalization
    \item On-the-fly data augmentation
    \item Patch-based 3D u-net
    \item Training using the Dice loss function for a pre-defined number of epochs
    \item Prediction on data from the validation set to gauge the performance on unseen data and to aid in hyperparameter tuning
\end{itemize}

When the model is applied to evaluation data, a postprocessing step concatenates black (background class) voxels to the output masks so that their dimensions agree with the original CT scans.

\subsection{Preprocessing}
Each of the 30 CT images was cut in half, splitting the left and right proximal femurs into separate images. The resulting CT scans that included the left proximal femur were then mirrored to the right side. The training/validation set effectively became 60 images of the right side hip/upper leg with in-plane resolution $512\times256$ voxels and 98-148 slices. The CT images were normalized such that all intensity values were linearly shifted and scaled from HU to the range $[0,1]$. Min-max normalization has the advantage over z-score normalization of preserving the scale of the data. Models using both normalization methods were implemented, however, there was no discernible difference between the two.

\subsection{Data Augmentation and Regularization}
\begin{figure}[t]%
\centering
\subfigure[Original]{%
\includegraphics[height=1.8in]{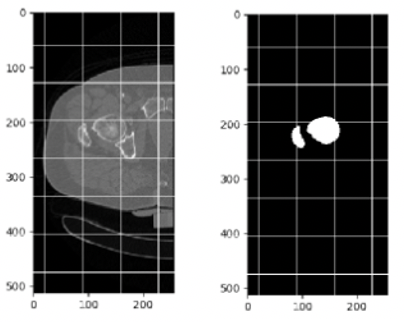}}%
\qquad
\subfigure[Deformed]{%
\includegraphics[height=1.8in]{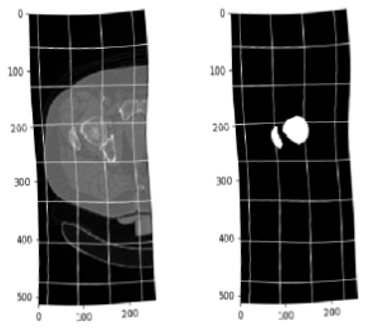}}%
\caption{The effect of elastic deformation on a 2D CT image and its mask.}
\label{elastic}
\end{figure}

Since obtaining manually segmented images is laborious and slow, the use of data augmentation is crucial for maximizing the efficiency of the training set. Data augmentation is used to teach the neural network invariance and robustness properties when a limited data set is available, thus artificially expanding the initial training set to avoid overfitting. These deformations can be simulated efficiently and aid the model in learning invariance between samples \citep{unet}. Here we applied both linear-spatial and intensity transformations (i.e., scaling, rotation and brightness) as well as elastic deformation (Figure~\ref{elastic}) to simulate the variability between patients' scans using the Batchgenerators package \citep{batchgenerators} within the Medical Image Segmentation with Convolutional Neural Networks (MIScnn) framework\footnote{MIScnn is an open-source Python library and intuitive API for medical image segmentation pipelines \citep{miscnn}.}. The exact parameter ranges implemented for our proposed model are given in Table \ref{params_dataaug}. 
\begin{table*}[t]
\centering
\caption{The data augmentation parameter ranges for the proposed model. Here, $\alpha$ denotes the scaling factor (controls the deformation intensity) and $\sigma$ denotes the smoothing factor (controls the displacement field smoothing) for the elastic deformation.}
     \sffamily \begin{tabularx}{1.0\textwidth}{ p{2cm}  p{2cm}  p{2cm} p{2cm} p{4cm}}\\
    \hline
   \textbf{} \hfill & \textbf{Brightness} \hfill & \textbf{Rotation (X,Y,Z)}  \hfill & \textbf{Scaling}  \hfill & \textbf{Elastic deformation} \\ \hline
    Parameter range & $(0.75,1.25)$ & $(-3^{\circ}, 3^{\circ})$ & $(0.95, 1.05)$ &  \begin{tabular}{@{}c@{}}$\alpha=(0,100)$ \\ $\sigma=(9,13)$\end{tabular}\\\hline
    \end{tabularx} \normalfont
\label{params_dataaug}
\end{table*}

Data augmentation, with random transformation parameters from the pre-defined ranges was performed on-the-fly for each image before it was forwarded into the neural network. Each of the data augmentation transformations had a 35\% likelihood of being applied to the image at hand, allowing the model to encounter a diverse set of images, thereby decreasing redundancy. 
For the proposed method, on-the-fly\footnote{On-the-fly data augmentation eliminates the need for excessive storage of augmented images by performing the augmentation prior to each optimization iteration.} data augmentation, in concert with parameter sharing \citep{lecun-boser} and batch normalization (BN) \citep{BN}, rendered the use of explicit regularization techniques unnecessary and even counterproductive.

\subsection{Model Architecture}
\begin{figure*}[t]%
\centering
\includegraphics[width=.9\textwidth]{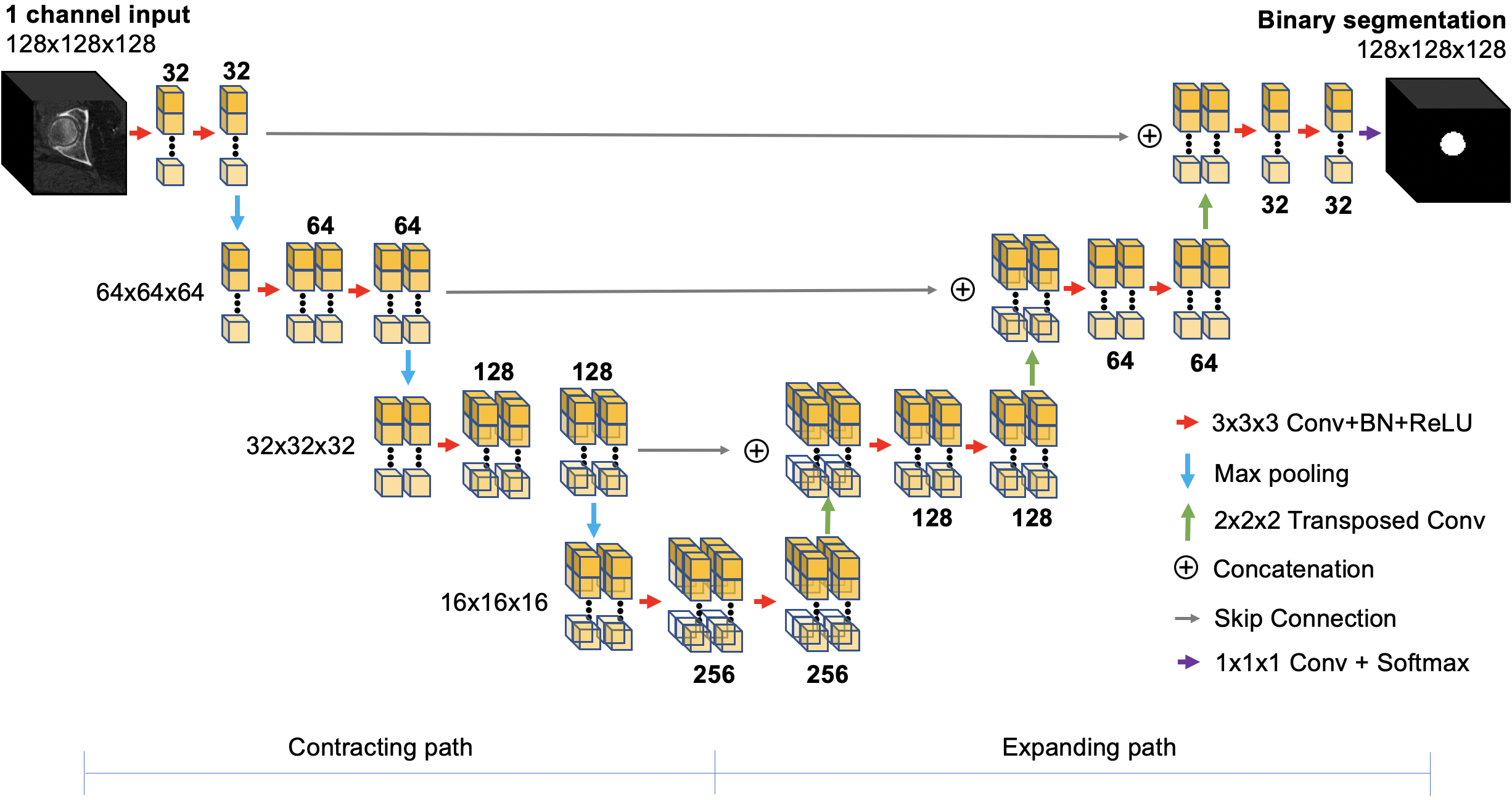}
\caption{A schematic of our proposed 3D u-net. The bold numbers at the corners represent the number of feature maps (channels) per layer. Here, convolution is abbreviated as \emph{conv} and rectified linear unit as \emph{ReLU.}}
\label{unet_standard_schematic}
\end{figure*}

 DNNs have prevailed as the state-of-the-art learning models for biomedical image segmentation, most notably the renowned \textit{u-net} \citep{unet}. This impactful and elegant network architecture, based on the FCN, addresses two main issues: namely, the ability to train a model from a very small data set and the ability to produce precise segmentations despite the former. A schematic of the proposed architecture is shown in Figure~\ref{unet_standard_schematic}. The u-net derives its name from the u-shape of the model architecture, consisting of a contracting (downsampling) path and an expanding (upsampling) path. The contracting path is the encoder and captures the context in the CT image by way of stacked convolutional and max pooling layers. The expanding path, on the other hand, is the decoder and allows for precise localization with the use of transposed convolutions. In the final layer of the network, a $1\times1\times1$ convolution is used to map the feature map to the number of classes. These outputs are of the same dimensions as the input volume and are converted to probabilistic segmentations of the foreground and background regions by applying a softmax layer voxel-wise. The voxels with a probability $>0.5$ belong to the foreground class (proximal femur) and the rest to the background class. 
The proposed neural network model architecture was implemented using the flexible MIScnn framework \citep{miscnn} in Python.

\subsection{Hyperparameter Selection}
A patch-based model, as opposed to analysis of the full image, was adopted in consideration of memory constraints and to exploit random cropping of patch volumes from the full images, further regularizing the model architecture. For the proposed u-net model, a patch volume of $128\times128\times128$ voxels with an overlap of $64\times64\times64$ voxels was forwarded to the network. This patch size is large enough to capture the entire femoral head, which is the most dynamic section of the proximal femur. Additionally, since the number 128 is readily divisible by two, we are left with integer values for patch dimensions after each use of max pooling.
\begin{table}[t]
\centering
\caption{Model parameters for the proposed neural network.}
     \sffamily \begin{tabularx}{1.0\columnwidth}{ p{6.5cm}  p{6.5cm} }
    \hline
   \textbf{Model parameter} \hfill & \textbf{Proposed value} \hfill \\ \hline
    Batch size & 2\\
    Patch size & $128\times128\times128$\\ 
    Layers & 4\\
    Feature map (highest resolution) & 32\\
    Initial learning rate & $1 \cdot 10^{-4}$\\ 
    Epochs & 300\\
    Iterations & 80\\
    Training duration (hrs.) & 12\\
    Adam optimizer \citep{adam} & $\beta_1 = 0.9$ and $\beta_2 = 0.999$\\\hline
    \end{tabularx} \normalfont
\label{params_standard_res}
\end{table}

A batch size of two, randomly cropped volumes of size $128 \times 128 \times 128$ appeared to be the optimal combination w.r.t. memory constraints. This combination consistently outperformed stochastic models with the same size patch volumes or larger, as well as outperforming models with larger batch sizes, which necessitated smaller patch volumes to avoid memory overload. When implementing a model with a batch size of one, the loss function fluctuates heavily since it is only considering one sample at a time. When the batch size was increased to four and a smaller patch size of $64\times64\times64$ was used, the model performance slightly decreased because of the limited context in each patch. A variety of combinations were tested to arrive at these conclusions. Our proposed model was tuned to the parameter values displayed in Table \ref{params_standard_res}.

\subsection{Training}
Our model was trained using a single Nvidia GeForce GTX 1080 Ti GPU for 300 epochs, which took roughly 12 hours. We randomly selected 30 CT images for training with corresponding manual segmentations of the left and right femur. Of these 60 proximal femurs, 54 were used for training and 6 were set aside to validate the performance of the model on unseen data. The number of slices was in the range of 98 to 148 slices. The ground truth annotations that comprised the training and validation sets are binary images identifying the voxels of the femur. 

\subsection{Postprocessing of Image Data}
The postprocessing step of the masks was twofold: Firstly, each mask was padded with black voxels that were cropped out during preprocessing. The segmentation predictions were hence restored to the original resolution of the ground truth segmentations ($512 \times 512$ voxels in-plane) and the same offset in the coordinate system. Secondly, the largest connected component had to be extracted to filter out noise in some of the segmentations outputted by our model. 

\section{Experiments and Results} \label{results}
To evaluate our segmentation method we conducted three experiments: a comparison with a state-of-the art femur segmentation approach using Sample I, an evaluation on Sample II, and an FE analysis to assess the viability of using our model as part of our hip fracture screening tool. 
\subsection{Evaluation Criteria}
We used two evaluation metrics to evaluate the accuracy and robustness of our segmentation predictions, the DSC, (as discussed in Section \ref{validation_and_loss_function} above) and the HD. While models seldom attempt to directly minimize the HD, this metric provides valuable insight into the performance of our model. This method quantifies the largest segmentation error by outputting the greatest distance from a point on the surface of the predicted segmentation mask to the closest point on the other surface of the ground truth segmentation mask. If $X$ and $Y$ are two non-empty subsets, the one-sided HD from $X$ to $Y$ is defined by the following equation:
\newcommand{\norm}[1]{\left\lVert#1\right\rVert}
\begin{equation} \label{eq:haus1}
    \begin{aligned}
    \tilde{\delta}_{H}(X,Y) = \max_{x \in X} \min_{y \in Y} \norm{x - y}_2.
    \end{aligned}
\end{equation}
Similarly, going from $Y$ to $X$ yields 
\begin{equation} \label{eq:haus2}
    \begin{aligned}
    \tilde{\delta}_{H}(Y,X) = \max_{y \in Y} \min_{x \in X} \norm{x - y}_2.
    \end{aligned}
\end{equation}
The bidirectional HD between these two sets is defined as:
\begin{equation} \label{eq:haus3}
    \begin{aligned}
    \delta_{H} = \max(\tilde{\delta}_{H}(X,Y), \tilde{\delta}_{H}(Y,X)).
    \end{aligned}
\end{equation}
The function $\tilde{\delta}_{H}(X,Y)$ finds the nearest point in $Y$ to each point in $X$, and selects the largest distance. The bi-directional HD measures the degree of mismatch between the two subsets by taking the the maximum value between the one-sided HDs, as shown in (\ref{eq:haus3}). 
It is common practice in biomedical image segmentation to use the \nth{95} percentile Hausdorff distance (HD95) in order to eliminate the influence of a small subset of outliers. 
\subsection{Comparison with the Graph-Cut Method}
\begin{figure}[t]%
\centering
\subfigure[DSC]{%
\label{dice_compare_CNN_GC_MAN}%
\includegraphics[height=2in]{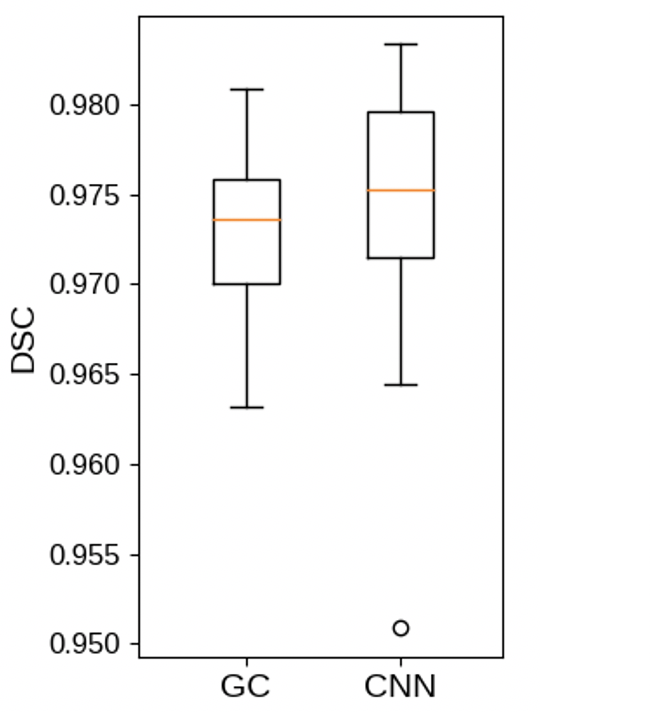}}%
\qquad
\subfigure[HD95]{%
\label{haus_compare_CNN_GC_MAN}%
\includegraphics[height=2in]{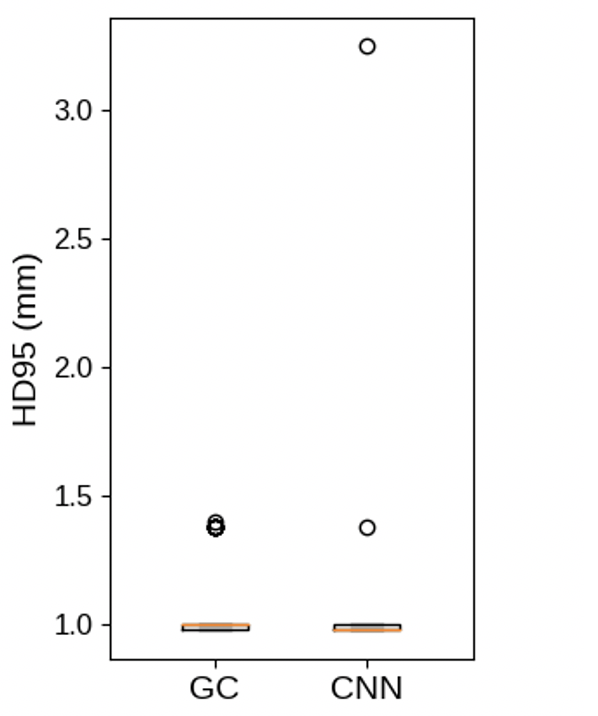}}%
\caption{A comparison between our method (CNN) and the graph-cut method (GC) on the same 24 CT image set (48 proximal femurs) of the left and right femurs (Sample I) validated on manual ground truth segmentations. Box plots (a) and (b) show the DSC and HD95, respectively, for the two methods.}
\end{figure}

A direct comparison was carried out between the proposed method and the graph-cut method by \citep{yves} to demonstrate the effective\-ness of our method on 24 unseen CT scans (Sample I). We computed the DSC and HD95 to quantitatively assess the accuracy of the two methods compared with ground truth manual segmentations (the current gold standard). As shown in Figures \ref{dice_compare_CNN_GC_MAN}, \ref{haus_compare_CNN_GC_MAN}, and Table \ref{graph-cut-table}, our method achieved a higher mean DSC score of $0.975\pm0.006$ and a lower HD95 of $1.04\pm0.33\mathrm{mm}$ than that of the graph-cut method ($0.973\pm0.005$ and $1.06\pm0.16\mathrm{mm}$, respectively). As displayed in the figures, there is one outlier in the CNN prediction. 
\begin{table*}[t]
\centering\setcellgapes{2pt}\makegapedcells \renewcommand\theadfont{\normalsize\bfseries}%
\caption{A comparison between the graph-cut method \citep{yves} and our method w.r.t. the DSC and HD95 validation metrics. Let $\sigma$ denote the standard deviation. We note that the results of the GC method were manually corrected for 47/48 proximal femurs. }
     \sffamily \begin{tabularx}{1.0\textwidth}{ p{3.5cm}  p{3cm}  p{4cm} p{3cm}}
    \hline
   \textbf{Method} \hfill & Mean DSC $\pm \sigma$ \hfill & Mean HD95 $\pm \sigma$ [mm] & Time [s]\\ \hline
    Graph-cut method & $0.973\pm0.005$ & $1.06\pm0.16$ & 120-300\\
    Proposed method & $0.975\pm0.006$ & $1.04\pm0.33$ & 9\\\hline
    \end{tabularx} \normalfont
\label{graph-cut-table}
\end{table*}

\begin{figure}[hbt!]%
\centering
\subfigure[]{%
\label{cyst_case}%
\includegraphics[height=3in]{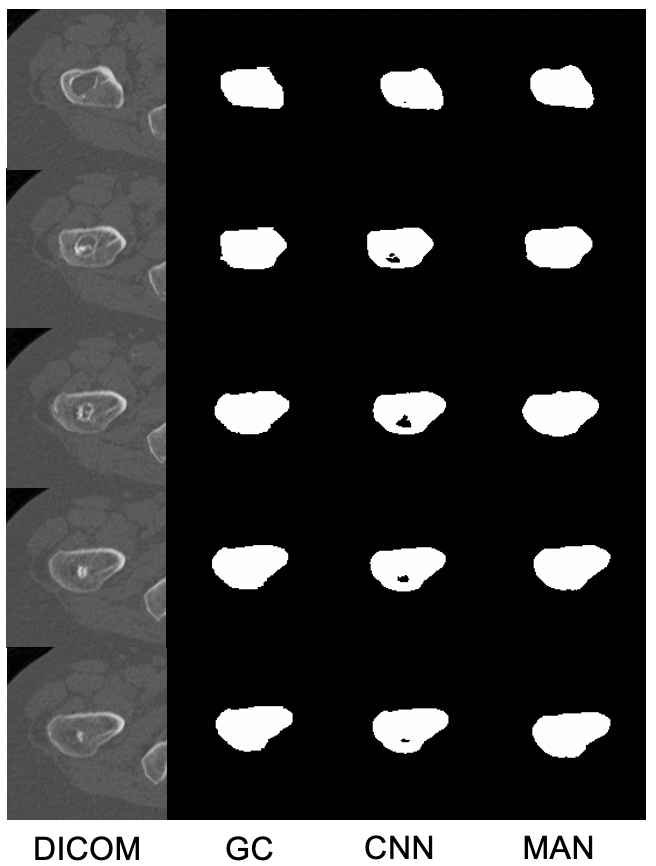}}%
\qquad
\subfigure[]{%
\label{normal_case}%
\includegraphics[height=3in]{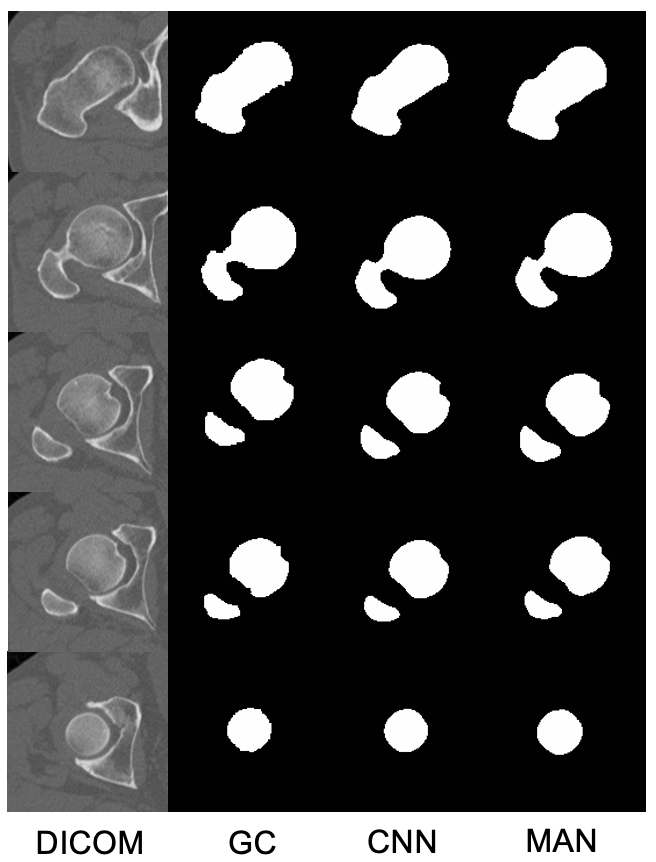}}%
\caption{A comparison between the original CT scan, the graph-cut method (GC), our method (CNN), and the manual ground truth segmentations (MAN) on five axial slices from two different patients in Sample I. Our method attempts to segment an artifact within the femoral shaft over the span of 16 slices (five shown) for a single case (a), however, it performs very well in all other cases, an example is shown in (b).}
\label{comparison_normal_cyst}
\end{figure}

The outlier in Figure~\ref{dice_compare_CNN_GC_MAN} around $\mathrm{DSC}=0.951$ corresponds with the outlier in Figure~\ref{haus_compare_CNN_GC_MAN} around $\mathrm{HD95}=3.25\mathrm{mm}$. Further investigation of the nature of the original CT scan revealed a possible cyst or the aftermath of intramedullary nailing to the femoral shaft (see Figure~\ref{cyst_case}). Since our method attempted to segment the structure inside of the bone, the DSC and HD95 metrics suffered moderately. This inadvertent labeling within the bone is, in part, a consequence of an absence of similar cases within the training set of the neural network. Data augmentation cannot be expected to simulate this type of variation if data of this kind are excluded from the training set. Figure~\ref{normal_case} shows the results on a subject in which both methods performed well.   
We note that this comparison is not completely fair in the sense that a manual operator has corrected all but one (47/48) of the proximal femur segmentations outputted by the graph-cut algorithm. Nevertheless, it shows that similar results are achieved in a small fraction of the time (only 11s on average for the proposed CNN as opposed to 2-5 minutes for the graph-cut method) and in a completely automated manner.

\subsection{Performance on Sample II of AGES}
We demonstrate the performance of our model on the aforementioned Sample II subset by evaluating it on 1147 previously unseen proximal femurs that have been segmented semi-automatically.
\begin{figure}[hbt]%
\centering
\subfigure[DSC]{%
\label{1155_a}%
\includegraphics[height=2in]{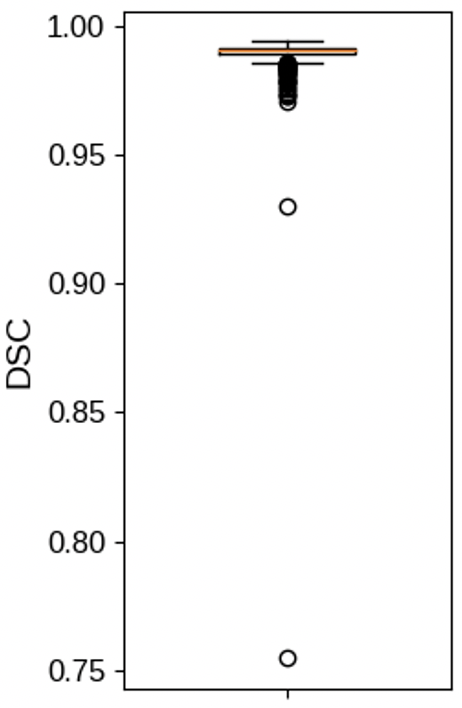}}%
\qquad
\subfigure[HD95]{%
\label{1155_b}%
\includegraphics[height=2in]{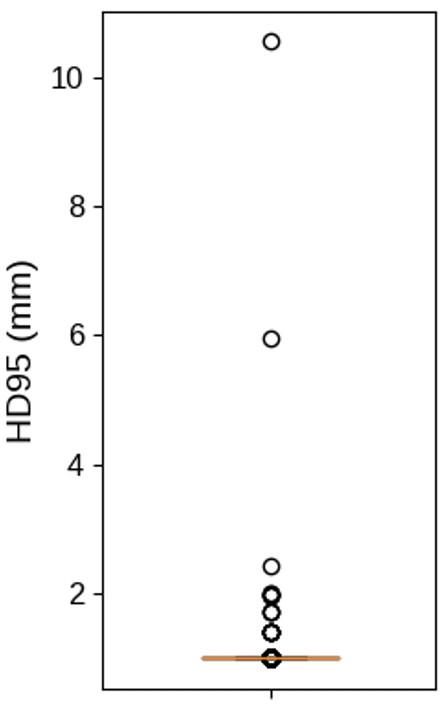}}%
\caption{The distribution of DSC and HD95 scores on Sample II.}
\label{1155_dice_haus}
\end{figure}
The box plots for both the DSC and HD95 scores are displayed in Figure~\ref{1155_dice_haus}. The mean DSC score was $0.990\pm0.008$ and the mean HD95 was $0.999\pm0.331\mathrm{mm}$. Only two data points had a DSC $<$ $0.97$ and a HD95 $>$ $2.4\mathrm{mm}$ (corresponding to the same two proximal femurs). The high average DSC, low HD95 value, and only two erroneous outliers out of a total of 1147 proximal femurs clearly reveal both the high accuracy and robustness of the proposed method. The time for each segmentation prediction averaged 11 seconds, which to our best knowledge is significantly faster than any current method, rendering our method viable for application to both large studies and clinical settings.

\begin{figure}[hbt]
\centerline{\includegraphics[width=\columnwidth]{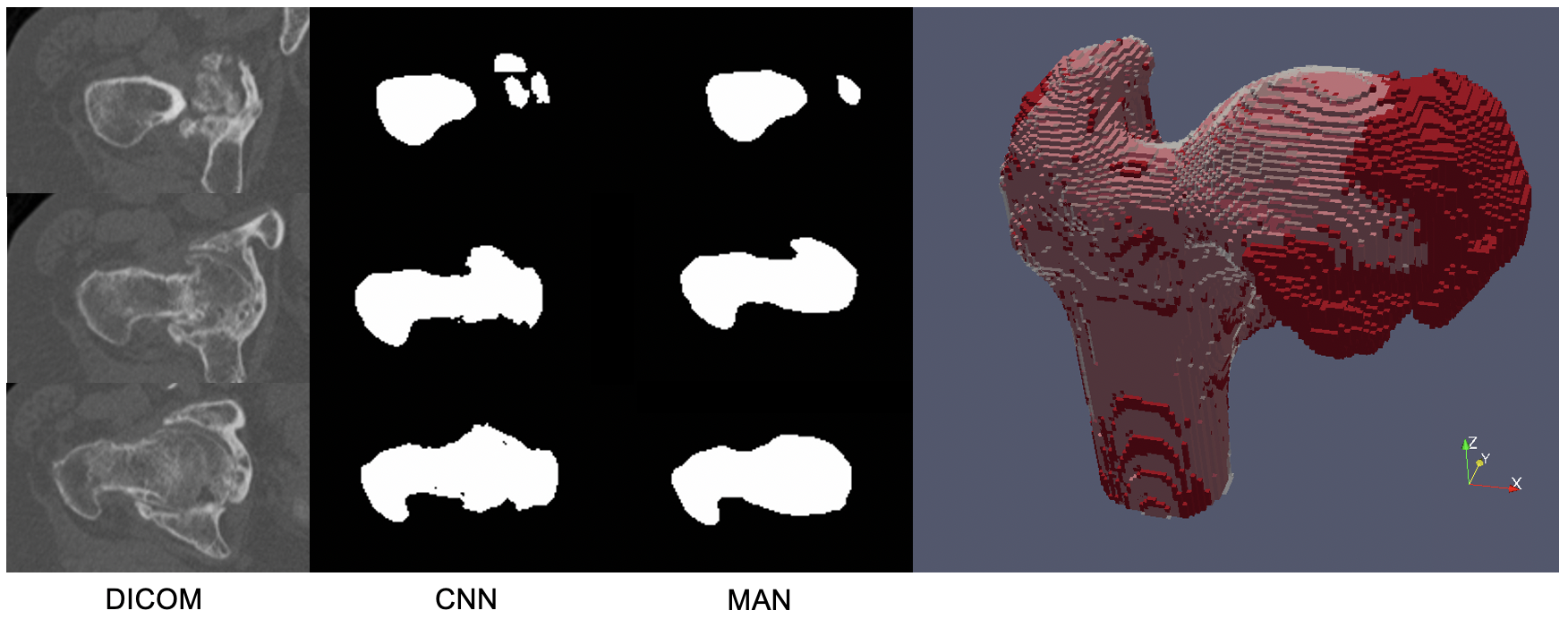}}
\caption{Three axial slices are displayed on the left hand side. The shorthand ``DICOM'' refers to the original CT scan, ``CNN'' refers to our method's segmentation, and ``MAN'' refers to the manually delineated ground truth segmentation. On the right, a 3D rendering of the erroneous segmentation prediction from our model (red) is overlaid with the ground truth segmentation (white).}
\label{93}
\end{figure}

\begin{figure}[hbt]
\centerline{\includegraphics[width=\columnwidth]{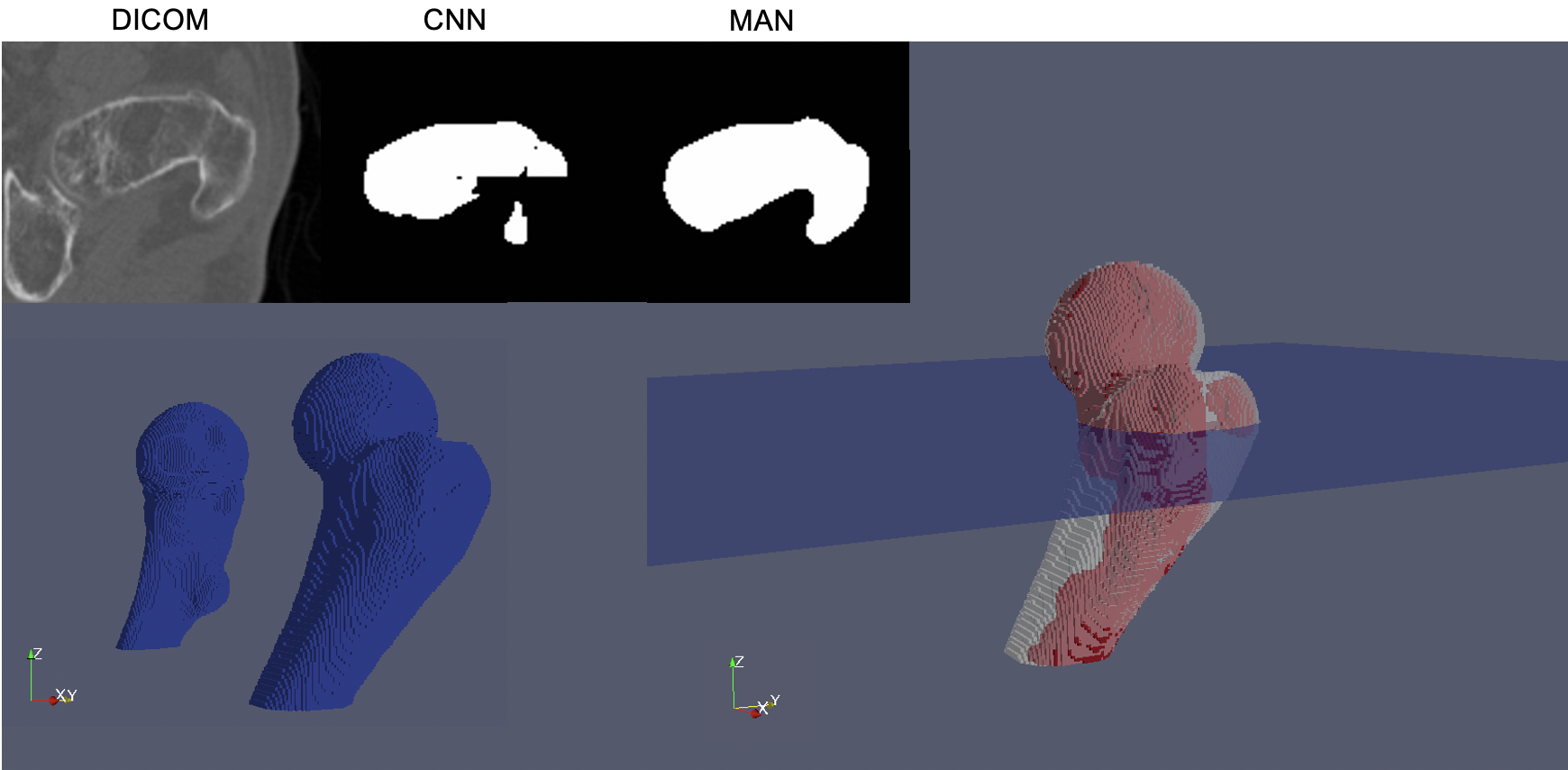}}
\caption{A single axial slice is displayed in the upper-left corner where the plane cuts the figure on the right. The left and right proximal femurs are shown on the bottom-left to illustrate the slant in the sagittal (YZ) plane. On the right, the erroneous segmentation prediction from our model (red) is overlaid with the ground truth segmentation (white).}
\label{75}
\end{figure}

One of the segmentation predictions from our model that received a slightly lower DSC score and higher HD95 score ($\mathrm{DSC} = 0.930$ and $\mathrm{HD95} = 5.94\mathrm{mm}$) on the right femur is shown in Figure~\ref{93}. The region around the head of the proximal femur was heavily over-segmented, as shown in the 3D rendering of Figure~\ref{93} (right side). This comes as no surprise considering how unclear the separation is between the femoral head and the acetabulum in the axial view of the CT image (Figure~\ref{93}, left). This is perhaps an indication of a birth defect or the result of a fractured bone that has since healed, however, any appraisal of the pathology without supplementary information on the subject is purely speculative. The other case that generated subpar results ($\mathrm{DSC}=0.755$ and $\mathrm{HD95}=10.6\mathrm{mm}$) is shown in Figure~\ref{75}. The left femur appears to be tilted in the sagittal plane, causing our model to output a poor, and even fragmented segmentation for some axial slices. The tilt could be the result of femoral anteversion (in-toeing), a lenient adherence to imaging protocol, or a multitude of other reasons. The implementation of bone registration preprocessing step to enforce spatial normalization could be a requisite tool in achieving acceptable segmentations for this phenomenon. 

\subsection{FE Pipeline Results}
\begin{figure}[h!]
\centerline{\includegraphics[width=\columnwidth]{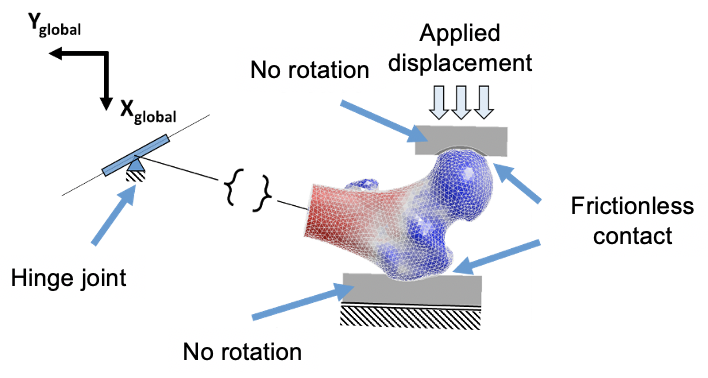}}
\caption{An FE Model of the proximal human femur (Image reprinted from Fleps et al. 2020, with permission).}
\label{FEM_model}
\end{figure}

In our last experiment we wanted to assess if our automated femur segmentation method could replace the manual segmentation approach currently being used in our FE pipeline for hip fracture predictions, giving way to a fully automated, end-to-end hip fracture screening process.
The FE models were based on the automated femur segmentations from the proposed method using an automated pipeline based on in-house Python scripts and a commercial preprocessor (Ansa 20.0.0; Beta CAE Systems, Switzerland). Models were solved using LS-Dyna (LS-Dyna v11.0, LS-Dyna, Livermore, CA, USA) and results postprocessed in Python. A detailed description of the modeling strategy was published in \citep{Ingmar} but is briefly described here for clarity and context. The proximal femurs were meshed with 10-node tetrahedral elements with an average mesh size of 3 mm. Heterogeneous literature-based non-linear material properties were assigned to the mesh based on CT gray scale values following a validated material mapping procedure \citep{Enns-bray, Fleps}. A femur loading alignment and boundary conditions representative of an unprotected fall to the side (10 adduction and 0 degrees internal rotation) were modeled. A schematic of the FE model of the proximal human femur is shown in Figure~\ref{FEM_model}. This femur modeling has shown improved hip fracture classification performance compared to aBMD in the AGES RS cohort  \citep{Ingmar}. Femurs were loaded until peak force was exceeded. Femoral strength was evaluated by recording the maximum force that the femur was able to withstand. The FEA-derived femoral strength, based on the semi-automated (manual segmentations from the graph-cut method \citep{yves}) and automated (our proposed method) approaches, was compared using the coefficient of determination ($\mathrm{R^2}$), root mean square error ($\mathrm{RMSE}$), mean absolute difference ($\mathrm{MAE}$), and the maximum difference (Table \ref{r_squared}). The data used were the predicted segmentations and ground truths from Sample II. 
\begin{table}[htb]
\centering\setcellgapes{2pt}\makegapedcells \renewcommand\theadfont{\normalsize\bfseries}%
\caption{The $\mathrm{R^2}$, $\mathrm{RMSE}$, $\mathrm{MAE}$, and maximum difference for the FEA-derived femoral strength values between the automated and manual methods (left femurs and right femurs).}
     \sffamily \begin{tabularx}{1.0\columnwidth}{ p{2.75cm}  p{1.75cm} p{2.5cm} p{2.25cm} p{3cm}}
    \hline
   \textbf{Proposed Method} \hfill & $\mathrm{R^2}$ \hfill &  $\mathrm{RMSE}$ [N]\hfill & $\mathrm{MAE}$ [\%]\hfill & Max Difference\hfill [\%]\\ \hline
    Left femur & $0.986$ & $212.2$ & $-2.14$ & $25.3$\\
    Right femur & $0.988$ & $177.0$ & $-1.86$ & $30.1$\\ \hline 
    \end{tabularx} \normalfont
\label{r_squared}
\end{table}

\begin{figure}[hbt]
\centerline{\includegraphics[width=\columnwidth]{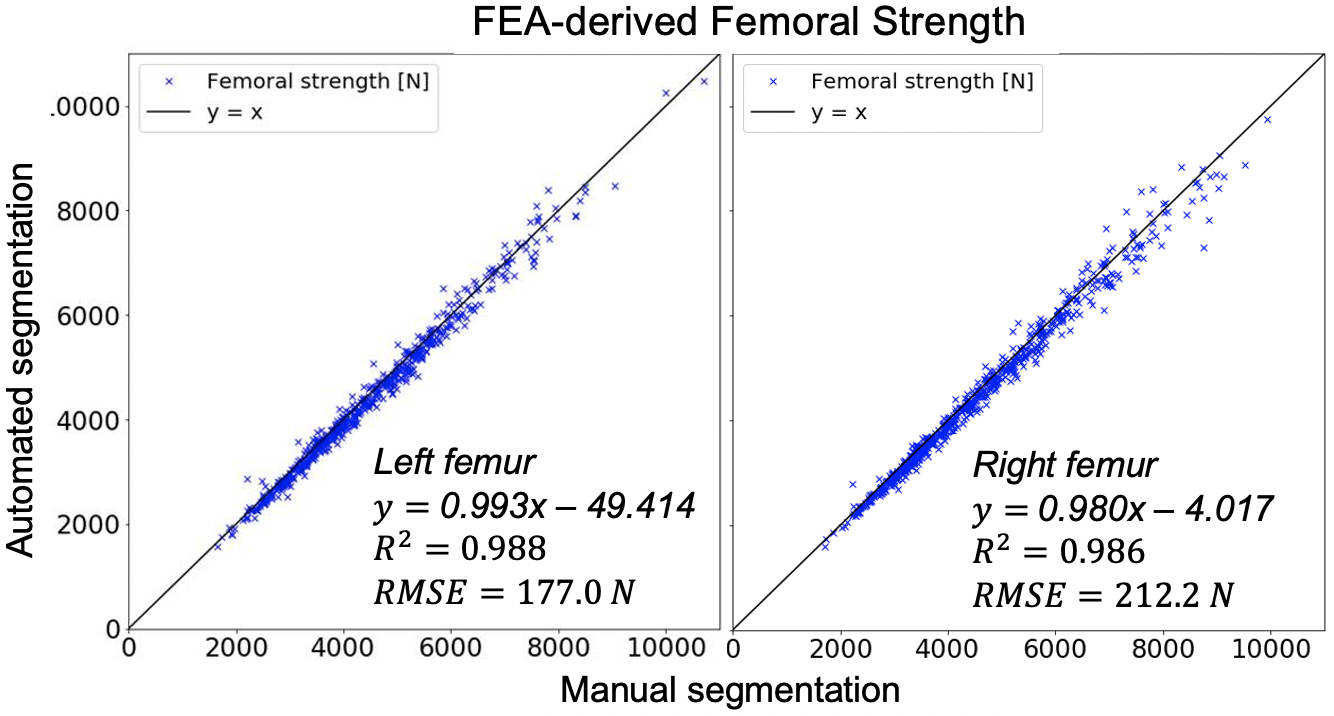}}
\caption{FEA-derived femoral strength based on the proposed femur segmentation method compared to femoral strength based on manual femur segmentations for left (left figure) and right (right figure) femurs.}
\label{FEM_correlation}
\end{figure}

Of the 611 subjects we segmented for the left femur, 593 simulation models were solved while 18 were exempted due to modeling errors (e.g., femurs that were not part of the cohort, data processing errors on the FE side, self-intersecting meshes, or the presence of extraneous volumes). Of these 593 models, 583 corresponding models based on the semi-automated segmentation were available to us. The predicted femoral strength values derived from the two segmentation methods were highly correlated (Figure~\ref{FEM_correlation}). Of the 576 subjects we segmented for the right femur, 562 simulation models were solved while 14 models were exempted due to modeling errors. Of these 562 models, 553 corresponding models based on the semi-automated segmentation were readily available (see Table \ref{model_numbers}). As displayed in Table \ref{r_squared}, similar results were achieved for both the left and right femurs, showing a very strong linear relationship between FEA-derived femoral strength from our fully automatic segmentations and from the semi-automatic segmentations. 
These results demonstrate that our method's segmentations are suitable for the FE pipeline and can be channeled through it in a robust manner.

\begin{table}[htb]
\centering
\caption{The number of segmented femurs, simulations run, and the number of femurs used to compare the automated method to the semi-automated method.}
     \sffamily \begin{tabularx}{1.0\columnwidth}{ p{8.5cm}  p{2.5cm}  p{2.5cm}}
    \hline
   \textbf{Femur set} \hfill & Left femur \hfill & Right femur \hfill\\ \hline
    Segmented femurs & 613 & 591\\
    Evaluated by pipeline & 593 & 562\\
    Used for the comparison to manual segmentation & 583 & 553\\\hline\hline
    \textbf{Excluded femurs} \hfill &  \hfill &  \hfill\\ \hline
    Corresponding left femur missing \hfill & - & 12\\
    Excluded due to FEA errors & 2 & 2\\
    Excluded due to segmentation or meshing errors & 16 & 13\\
    Not part of cohort & 2 & 2\\\hline
    \end{tabularx} \normalfont
\label{model_numbers}
\end{table}

\section{Discussion}
The aim of this work was to develop a fully automated neural network for proximal femur segmentation from CT images for application to an existing FE pipeline for hip fracture risk prediction. We demonstrated that our model's performance in terms of the DSC and HD95 is comparable to that of one of the previous best methods \citep{yves}, yet significantly faster and without a human interaction. We subsequently presented our model's evaluation performance on 1147 unseen proximal femurs from the AGES-RS Sample II cohort \citep{ages}, achieving a mean DSC of $0.990\pm0.008$ and a mean HD95 of $0.999\pm0.331\mathrm{mm}$. Lastly, we demonstrated a $\mathrm{R^2}$ value of approximately $0.987$ between FEA-derived femoral strength values based on our method's segmentations and based on manual segmentations. 

The comparison with the graph-cut method \citep{yves} demonstrates our model's superior nature in terms of accuracy and robustness, despite not having a trained human operator to correct unacceptable segmentations ad libitum. Not only does our method output segmentation predictions an order of magnitude faster than the graph-cut method, but additionally relieves our future end-users (e.g., health care practitioners) from the monetary cost of hiring a trained specialist to perform the corrections. This is one of the significant hurdles that prior methods have struggled to surmount.

The results from Sample II demonstrate our model's undeniable accuracy and robustness, allowing us to process even larger cohorts in the near future.
With regard to the very few problematic cases encountered in Sample II, we speculate which measures are justifiable to take in order to further increase the robustness of our model. For the CT scans in which the proximal femur appears to slant, as in Figure~\ref{75}, we can argue that the use of registration to a common coordinate system would improve our model's prediction. Registration will eliminate the need to capture the variability within the data set for some of the most extreme cases with data augmentation. The use of such aggressive augmentation parameters is a futile pursuit that severely hampers the overall performance of the model, considering its sensitivity to radical transformations. By spatially transforming a source image to align with a target image, representing the mean shape constructed from a statistical atlas of healthy patient CT images, we enforce spatial normalization to the source image. We must, however, ask ourselves whether these preprocessing measures are worth the added effort considering how infrequently we encounter such anomalies. It is reasonable to assume that in clinical practice, physicians would immediately flag any patient scans that deviate significantly from the mean and would not be good candidates for our hip fracture screening tool. If our evaluation set is any indicator of the prevalence of anomalies in the general population, then this would amount to a negligible number of patients who could not be screened with our method. We note that in order to apply our model to CT images from a different scanner, the model would likely have to be trained on a set of images from that CT scanner.

In the last part of the Results section, we fed our model's segmentation predictions to the FE pipeline. The strong $\mathrm{R^2}$ values between FEA-derived femoral strength values based on our model's segmentations and manual ones, demonstrate our method's ability to reliably produce segmentations that can be processed by the FE pipeline with very similar predicted femoral strength. 

The primary limitations of this research are twofold: firstly, we have yet to demonstrate our solution’s ability to perform well on cross-cultural data, that is, on CT images beyond the Icelandic elderly as well as its performance on scans from different scanner manufacturers. If the performance of our model turns out to be unsatisfactory on other cohorts, then a possible solution would be to re-train the model with either a mix of scans from multiple populations or exclusively train on images from the cohort at hand. However, the desired outcome would be a segmentation tool that can be applied to all proximal femur scans independent of population and scanner manufacturer. The second limitation of our model is its performance on heavily deformed proximal femurs. There is an inherit trade-off between general segmentation performance and variability within the training set. That is, if we bias the training set with too many deformed bones, we will compromise the general performance on the validation set. As a result, we justify the exclusion of acutely deformed bones in the training set in order to improve performance on bones that do not deviate drastically from the mean.

In summary, our method has addressed the most pressing hindrances that have impeded prior methods; our method does not require a trained operator to make ad hoc corrections to unsatisfactory segmentations, the robustness of the model would justify the radiation exposure to the patient, and the processing time of each segmentation is well within reasonable bounds for clinical viability. More importantly, this demonstrates that with the proposed segmentation method, the hip fracture risk assessment can now be performed in a fully automated manner. The next step in the ongoing development of the hip fracture screening tool is applying the model to a much larger cohort of the AGES-RS. 

\section{Conclusion}
Here we introduced a fully automated, accurate, robust, and fast segmentation method for segmenting the proximal femur from CT images. The mean DSC was $0.990\pm0.008$ and mean HD95 was $0.999\pm0.331\mathrm{mm}$ when evaluated on 1147 manually segmented femurs. The proposed method is superior to preceding methods in terms of previously reported numbers of DSC and HD95 metrics and, most importantly, does not require any manual interaction. In addition, each segmentation prediction can be generated, on average, in 11 seconds instead of the many minutes it takes some other approaches. We will conduct a more extensive evaluation on a larger cohort and, in turn, integrate the method into our existing FE pipeline, bringing it one step closer to becoming a clinically viable option for screening at-risk patients for hip fracture susceptibility.

\section{Acknowledgements}
This research was supported in part by the RANNIS Icelandic student innovation fund, Iceland, and the Strategic Focus Area ``Personalized Health and Related Technologies'' of the ETH Domain, ETH Zurich, Switzerland [grant numbers 2018-430, 2018-325].

\section{Disclosure Statement}
The authors declare that they have no known competing financial interests or personal relationships 
that could have appeared to influence the work reported in this paper.

\bibliographystyle{tfcse}
\bibliography{interactcsesample}

\end{document}